\documentclass[technote,10pt]{IEEEtran}
\usepackage{cite}
\usepackage{color}
\usepackage{threeparttable}
\usepackage{graphicx}
\usepackage{picinpar}
\usepackage[cmex10]{amsmath}
\usepackage{amsmath,amsfonts,amssymb}
\usepackage{subfig}
\usepackage{changepage}
\usepackage{algorithm}
\usepackage{algpseudocode}
\usepackage{stfloats}
\usepackage{bm}
\usepackage{enumerate}

\usepackage{booktabs}
\usepackage{multirow}
\usepackage{etoolbox}
\usepackage{makecell}
\usepackage{xcolor}

\usepackage{stfloats}

\begin{document}
\newtheorem{lemma}{Lemma}
\newtheorem{corol}{Corollary}
\newtheorem{theorem}{Theorem}
\newtheorem{proposition}{Proposition}
\newtheorem{definition}{Definition}
\newcommand{\e}{\begin{equation}}
\newcommand{\ee}{\end{equation}}
\newcommand{\eqn}{\begin{eqnarray}}
\newcommand{\eeqn}{\end{eqnarray}}
\newenvironment{shrinkeq}[1]
{ \bgroup
\addtolength\abovedisplayshortskip{#1}
\addtolength\abovedisplayskip{#1}
\addtolength\belowdisplayshortskip{#1}
\addtolength\belowdisplayskip{#1}}
{\egroup\ignorespacesafterend}
\title{Compressive Sensing Based Channel Estimation for Millimeter-Wave Full-Dimensional MIMO with Lens-Array}

\author{Ziwei Wan, Zhen Gao, Byonghyo Shim,~\IEEEmembership{Senior Member,~IEEE},
Kai Yang, Guoqiang Mao,~\IEEEmembership{Fellow,~IEEE},
and Mohamed-Slim Alouini,~\IEEEmembership{Fellow,~IEEE}
\thanks{Z. Wan, Z. Gao, and K. Yang are with School of Information and Electronics, Beijing Institute of Technology, Beijing 100081, China.
Z. Gao is also with the Advanced Research Institute of Multidisciplinary Science, Beijing Institute of Technology, Beijing 100081, China (e-mail: gaozhen16@bit.edu.cn).

B. Shim is with Institute of New Media and Communications School of Electrical and Computer Engineering, Seoul National University, Seoul, Korea.

G. Mao is with School of Electrical and Data Engineering, The University of Technology
Sydney, Australia.}
\thanks{M.-S. Alouini is with the Electrical Engineering Program, Division of Physical
 Sciences and Engineering, King Abdullah University of Science and Technology (KAUST), Thuwal, Makkah Province, Saudi Arabia.}
}

\maketitle
\begin{abstract}
Channel estimation (CE) for millimeter-wave (mmWave) lens-array suffers from prohibitive training overhead, whereas the state-of-the-art solutions require an extra complicated radio frequency phase shift network. By contrast, lens-array using antenna switching network (ASN) simplifies the hardware, but the associated CE is a challenging task due to the constraint imposed by ASN. This paper proposes a compressive sensing (CS)-based CE solution for full-dimensional (FD) lens-array, where the mmWave channel sparsity is exploited. {Specifically, we first propose an approach of pilot training under the more severe haraware constraint imposed by ASN, and formulate the associated CE of lens-array as a CS problem. Then, a redundant dictionary is tailored for FD lens-array to combat the power leakage caused by the continuous angles of multipath components. Further, we design the baseband pilot signals to minimize the total mutual coherence of the measurement matrix based on CS theory for more reliable CE performance.} Our solution provides a framework for applying CS techniques to lens-array using simple and practical ASN. Simulation results demonstrate the effectiveness of the proposed scheme.
\end{abstract}
 \vspace*{-1mm}
 \vspace*{-2.0mm}
\IEEEpeerreviewmaketitle
\vspace*{-2mm}
\section{Introduction}

{Millimeter-wave (mmWave) is a key enabling technology for 5G and beyond \cite{Gao_TSP}, and its applications to vehicular communications have attracted significant attention in recent years \cite{Vehi}. Multiple-input multiple-output (MIMO) system with lens-array is a cost-efficient way to facilitate mmWave communications \cite{PDM, peak, FD, DC, DL, image, IA}.} By exploiting the energy-focusing property of large-aperture lens and small number of radio frequency (RF) chains, system can be implemented by the simple antenna switching network (ASN) instead of the bulky phase shifter network (PSN). Main benefit of this approach is that the spatial multiplexing can be achieved by using lens-arrays with reduced power consumption and hardware cost \cite{PDM, peak, FD}.

However, major challenge of this approach is that we need to estimate the high-dimensional channels from a limited number of RF chains \cite{PDM, peak, FD, DC, DL, image, IA}. In \cite{PDM}, the lens-based approach has been proposed. In this scheme, MIMO channel is divided into the multiple single-input single-output channels, assuming that angles of arrival (AoAs) and departure (AoDs) are separated sufficiently. On that basis, a channel estimation (CE) scheme for the linear lens-array has been proposed in \cite{peak}. In \cite{FD}, the idea in \cite{peak} has been further extended to the problem to estimate the channels between base station (BS) using full-dimensional (FD) lens-array and multiple users with analog precoding. However, it has been pointed out in \cite{IA} that the residual interference from different paths still exists, resulting in the degradation of performance. To address the problem, a beam selection scheme for the single-antenna users has been proposed. In \cite{DC, DL}, more sophisticated approaches to estimate the channels between FD lens-array and users with one or multiple single-antenna has been proposed. To support the multi-antenna users with analog precoding, a CE scheme utilizing the image reconstruction technique has been proposed \cite{image}. Drawback of the approaches in \cite{peak,FD} is that they require a complicated ASN, where one RF chain needs to activate all transmit antennas. Moreover, solutions in \cite{DC, DL, image} require an extra complicated PSN (see Fig. 1(a)), causing insertion loss, power consumption, and also extra hardware cost.

In this paper, we propose a compressive sensing (CS)-based CE technique for mmWave FD-MIMO with lens-array. In this scheme, we use a low-cost and energy-saving ASN where each RF chain is activating at most one antenna (see Fig. 1(b)). {First, we propose a framework of pilot training taking into account the constraint imposed by ASN. Then, we design a redundant dictionary tailored for FD lens-array to combat the power leakage caused by continuous AoAs/AoDs. Moreover, to minimize the {\it total mutual coherence} of the measurement matrix \cite{BShim,22,Elad,Gao_CS}, we design the transmit/receive pilot signals in the baseband (BB) part.} Simulations are conducted to demonstrate the effectiveness of the proposed scheme over the conventional approaches.

{Our contributions are summarized as follows.
\begin{itemize}
\item{We propose a CS-based CE approach that takes into account the constraint imposed by ASN. This is in contrast to the existing CS-based solutions in \cite{DC, DL, image} where a randomized PSN is employed to generate pilot signals, so that entries of the measurement matrix are independent identical distributed (i.i.d.) with good restricted isometry property (RIP), at the cost of complicated RF hardware.}

\item{To combat the power leakage in the angular-domain sparse CE for MIMO systems, we consider the unique antenna structure of lens-array and design a redundant dictionary tailored for FD lens-array to sparsify the channel and improve the sparse CE performance. {\it To the best of our knowledge, this is the first trial to design a redundant dictionary for the FD lens-array}.}

\item{We design the BB pilot signals for further improvement of CE performance. The state-of-the-art pilot design in \cite{DC, DL, image, Gao} depends on the randomized PSN to design the measurement matrix according to the RIP. However, these solutions are no longer applicable for lens-array with simple ASN, and RIP-based pilot design is very difficult in practical scenario \cite{BShim}. Hence, we design the pilot under a more tractable {\it total mutual coherence minimization} criterion \cite{BShim,22,Elad,Gao_CS}, whereby the closed-form solution to optimize the BB pilot signals can be derived.}
\end{itemize}}

\textit{Notations}:
Vectors and matrices are denoted by lower- and upper-case boldface letters, respectively.
${\left(  \cdot  \right)\!^*}$, ${\left(  \cdot  \right)\!^T}$, ${\left( \cdot \right)\!^H}$, and ${\mathop{\rm Tr}\nolimits} \left(  \cdot  \right)$ denote the conjugate, transpose, conjugate transpose and trace of a matrix, respectively.
$\mathbb{C}$ and $\mathbb{Z}$ are the sets of complex numbers and integers, respectively. $\mathcal{C}\mathcal{N}$ denotes the complex Gaussian distribution.
 ${[ \cdot ]_{i}}$ and ${[ \cdot ]_{i,j}}$ represent the ${i}$-th element of a vector and ${i}$-th row, ${j}$-th column element of a matrix, respectively.
${\bf{I}}_N$ represents the ${N \!\times \!N}$ identity matrix.
$\left\lfloor  \cdot  \right\rfloor $ and $\left\lceil  \cdot  \right\rceil$ denote the flooring function and ceiling function, respectively.
The ``sinc'' function is defined by ${\rm{sinc}} \left( x \right) \buildrel \Delta \over = {\rm{sin}} \left( \pi  x \right) / \left( \pi  x \right)$.
${\left\| {\cdot} \right\|_2}$, ${\left\| {\cdot} \right\|_{F}}$, and ${\rm{diag(}} \cdot {\rm{)}}$ represent the ${\ell _2}$-norm, Frobenius norm, and (block) diagonalization, respectively.
$ \otimes $ is the Kronecker product and ${\rm{vec(}} \cdot {\rm{)}}$ is the vectorization operation according to the columns of the matrix.

\begin{figure}[t]
	\captionsetup{font={footnotesize, color = {black}}, name = {Fig.}, labelsep = period}
	\label{fig:model}
	\centering
	\includegraphics[width=1\linewidth]{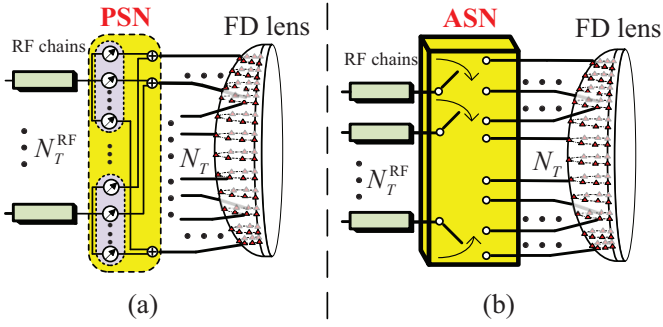}
	\caption{Diagrams of mmWave FD-MIMO with lens-array and (a) bulky full-connected PSN used in \cite{DC, DL, image}; (b) proposed simple ASN.}
\end{figure}

\vspace*{-2mm}
\section{System Model}
{We consider a mmWave FD-MIMO system with lens-arrays at both the transmitter and the receiver. In this work, we employ a simple and practical ASN, where each RF chain can activate at most one antenna, as shown in Fig. 1(b). Compared to the full-connected PSN (see Fig. 1(a)) \cite{DC, DL, image}, the proposed ASN is simple to implement and also power-efficient so that it is a more appealing model for MIMO communication under the limited RF power resource \cite{yangkai}.}
\color{black}
On the focal surface of electromagnetic (EM) lens, we use the antenna distribution for FD angular coverage proposed in \cite{FD}. In this model, the total number of transmit antennas is given by\footnote{Our distribution is slightly different from that in \cite{FD} to make $N_T$ be even without degrading the spatial resolution of lens-array for ease of analysis.}
\begin{equation}
\label{equ:Nt}
{N_T} = \sum\nolimits_{n = - \left\lfloor {\tilde D_T^v } \right\rfloor + \frac{1}{2}}^{\left\lfloor {\tilde D_T^v } \right\rfloor - \frac{1}{2}} {(2\left\lfloor {\tilde D_T^h\cos ({\rm{arcsin(}}\frac{n}{{\tilde D_T^v}}{\rm{)}})} \right\rfloor  + 1)} \text{,}
\end{equation}
where $\tilde D_{T}^h$ and $\tilde D_{T}^v$ are the normalized apertures of the transmit FD lens in the horizontal and vertical dimensions, respectively.
Similarly, for the normalized apertures $\tilde D_{R}^h \times \tilde D_{R}^v$ of receive FD lens, the number of receive antennas $N_R$ can also be calculated using (\ref{equ:Nt}).
The numbers of RF chains at the transmitter and the receiver are denoted by $N_{T}^{{\rm{RF}}}$ and $N_{R}^{{\rm{RF}}}$, respectively. Moreover, the associated channel matrix $\mathbf{H}\in {{\mathbb{C}}^{{{N}_{R}}\times {{N}_{T}}}}$ can be modeled as \cite{image}
\begin{shrinkeq}{-1ex}
\begin{align}
\hspace{-4.6mm}
{\bf{H}} = ({\sqrt {{N_T}{N_R}}}/{\sqrt L })\sum\limits_{l = 1}^L {{g _l}{{\bf{a}}_R}(\theta _R^l,\varphi _R^l)} {\kern 1pt} {\bf{a}}_T^H(\theta _T^l,\varphi _T^l) \text{,}
\end{align}
\end{shrinkeq}
where $L$ is the number of multipath components, ${{g }_{l}}$ is the complex gain corresponding to the $l$-th path, $\theta _{T}^{l}$ ($\theta _{R}^{l}$) and $\varphi _{T}^{l}$ ($\varphi _{R}^{l}$) are the vertical angle and horizontal angle of AoD (AoA) of the $l$-th path, respectively, ${\sqrt {{N_T}{N_R}} } / {\sqrt L }$ is the normalization factor, and ${{\mathbf{a}}_{T}}\in {{\mathbb{C}}^{{{N}_{T}}\times 1}}$ and ${{\mathbf{a}}_{R}}\in {{\mathbb{C}}^{{{N}_{R}}\times 1}}$ are the steering vectors of lens-arrays at the transmitter and receiver, respectively. The steering vectors for lens-arrays are different from those for phased-arrays. Taking the transmitter for instance, the steering vector can be expressed as
\begin{shrinkeq}{-1ex}
\begin{align}
\label{sinc}
\begin{split}
\hspace{-4.6mm}
{{[{{\mathbf{a}}_{T}}(\theta _T^l,\varphi _T^l)]}_{n}} & = \gamma \text{sinc }\!\![\!\!\text{ }{{\tilde{D}}_T^{v}}(\sin \alpha _T^n -\sin \theta _T^l)]\\
 \times \text{sinc } & \!\![\!\!\text{ }{{\tilde{D}}_T^{h}}(\sin \beta _T^n \cos \alpha _T^n -\cos \theta _T^l \sin \varphi _T^l)] \text{,}
\end{split}
\end{align}
\end{shrinkeq}
where (${{\alpha }_T^{n}}$,$\beta _T^n$) is the angular coordinate of the $n$-th antenna dependent on the location of the antenna on the focal surface \cite{FD} and $\gamma$ is a normalization factor guaranteeing $\left\|{{\mathbf{a}}_{T}}(\theta _T^l,\varphi _T^l) \right\|_{2}^{2}=1$.

\section{Proposed Channel Estimation Technique}
\subsection{Proposed CS Approach Based on Pilot Training}
\begin{figure}[t]
	 \captionsetup{font={footnotesize}, name = {Fig.}, labelsep = period}
     \centering
     \includegraphics[width=8.8cm, keepaspectratio]%
     {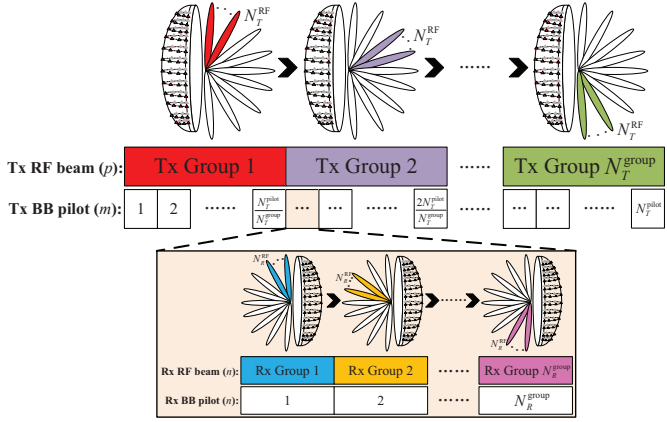}
     \caption{Illustration of proposed pilot training scheme using directional beams.}
     \label{fig:compressed}
\end{figure}

{Considering the hardware property of lens-array with ASN, we design a pilot training framework based on CS.} Specifically, the transmit pilot signal in the $m$-th time block ${{\mathbf{s}}_{m}}\in {{\mathbb{C}}^{{{N}_{T}}\times 1}}(1 \le m \le N_{T}^{\rm{pilot}})$ can be expressed as a product of the RF part $\mathbf{F}_{\rm{RF}}^{p}\in {{\mathbb{C}}^{{{N}_{T}}\times N_{T}^{\rm{RF}}}}$ and the BB part $\mathbf{f}_{\rm{BB}}^{m}\in {{\mathbb{C}}^{N_{T}^{\rm{RF}}\times 1}}$ as
\begin{shrinkeq}{-1ex}
\begin{align}
\begin{array}{*{20}{l}}
\hspace{-2mm}
{{\mathbf{s}}_m} = {\bf{F}}_{{\rm{RF}}}^p{\bf{f}}_{{\rm{BB}}}^m \text{.}
\end{array}
\end{align}
\end{shrinkeq}
{We assume $N_{T}^{{\rm{group}}} = {N_{T}}/N_{T}^{{\rm{RF}}} \in \mathbb{Z}$, $N_{T}^{{\rm{pilot}}} / N_{T}^{{\rm{group}}} \in \mathbb{Z}$ without loss of generality, and $p = \left\lceil {mN_T^{{\rm{group}}}/N_T^{{\rm{pilot}}}} \right\rceil \in \{ 1,2,...,N_T^{{\rm{group}}} \}$. This implies that every $N_{T}^{{\rm{pilot}}} / N_{T}^{{\rm{group}}}$ successive transmit BB pilot signals ${\bf{f}}_{{\rm{BB}}}^m$ will share the same transmit RF pilot signal ${\bf{F}}_{{\rm{RF}}}^p$.}
In our scheme, $N_T^{\rm{RF}}$ antennas are simultaneously activated as a transmit (Tx) group to form $N_T^{\rm{RF}}$ directional transmit beams, as illustrated in Fig. \ref{fig:compressed}.

{At the receiver, we assume that $N_{R}^{{\rm{group}}} = {N_{R}}/N_{R}^{{\rm{RF}}} \in \mathbb{Z}$ and all $N_{R}^{\rm{RF}}$ RF chains are used. Thus, each time block can be divided into $N_R^{{\rm{group}}}$ equal-length time slots} (see Fig. \ref{fig:compressed}). The received signal in the $n$-th time slot $(1 \le n \le N_{R}^{\rm{group}})$ from the $m$-th time block can be expressed as
\begin{shrinkeq}{-1ex}
\begin{align}
\begin{split}
\hspace{-4.6mm}
{{\bf{y}}_{n,m}} &  = {({\bf{W}}_{{\rm{RF}}}^n{\bf{W}}_{{\rm{BB}}}^n)^H} ( {\bf{H}}{{\bf{s}}_m} + {{\bf{n}}_{n,m}} ) \\
 & = {({\bf{W}}_{{\rm{RF}}}^n{\bf{W}}_{{\rm{BB}}}^n)^H}{\bf{H}}{{\bf{s}}_m} + {{{\bf{\bar n}}}_{n,m}} \text{,}
\end{split}
\end{align}
\end{shrinkeq}
where $\mathbf{W}_{\rm{RF}}^{n}\in {{\mathbb{C}}^{{{N}_{R}}\times N_{R}^{\rm{RF}}}}$ and $\mathbf{W}_{\rm{BB}}^{n}\in {{\mathbb{C}}^{N_{R}^{\rm{RF}}\times N_{R}^{\rm{RF}}}}$ are the RF and BB parts of the receive pilot signals, respectively, ${{\mathbf{n}}_{n,m}}\sim{\ }\mathcal{C}\mathcal{N}(0,{{\sigma _{n}^2\mathbf{I}}_{{{N}_{R}}}})$ is the additive white Gaussian noise vector, and ${{{\bf{\bar n}}}_{n,m}} = {({\bf{W}}_{{\rm{RF}}}^n{\bf{W}}_{{\rm{BB}}}^n)^H}{{\bf{n}}_{n,m}} \in {{\mathbb{C}}^{N_R^{{\rm{RF}}} \times 1}}$. Similar to the transmitter, $N_R^{\rm{RF}}$ antennas are activated as a receive (Rx) group and $N_R^{\rm{RF}}$ directional receive beams are formed. For $N_{R}^{\rm{group}}$ successive time slots at the receiver, we obtain ${{\mathbf{y}}_{m}}\in {{\mathbb{C}}^{{{N}_{R}}\times 1}}$ by collecting ${\rm{\{ }}{{\bf{y}}_{n,m}}{\rm{\} }}_{n = 1}^{N_R^{{\rm{group}}}}$. That is,
\begin{shrinkeq}{-1ex}
\begin{align}
\hspace{-4.6mm}
{{\mathbf{y}}_{m}}={{(\mathbf{W}_{\rm{RF}}^{{}}\mathbf{W}_{\rm{BB}}^{{}})}^{H}}\mathbf{H}{{\mathbf{s}}_{m}}+{{\mathbf{\bar{n}}}_{m}} \text{,}
\end{align}
\end{shrinkeq}
where ${{\bf{y}}_m} = {[{\bf{y}}_{1,m}^T,...,{\bf{y}}_{N_R^{{\rm{group}}},m}^T]^T} \in {{\mathbb{C}}^{{{N}_{R}} \times 1}}$,
${{\bf{\bar n}}_m} = [{\bf{\bar n}}_{1,m}^T,$ $...,{\bf{\bar n}}_{N_R^{{\rm{group}}},m}^T]^T \in {{\mathbb{C}}^{{{N}_{R}} \times 1}}$,
$\mathbf{W}_{\rm{RF}}^{{}} = [\mathbf{W}_{\rm{RF}}^{1},...,\mathbf{W}_{\rm{RF}}^{N_{R}^{\rm{group}}}]\in {{\mathbb{C}}^{{{N}_{R}}\times {{N}_{R}}}}$, and
$\mathbf{W}_{\rm{BB}}^{{}} = \text{diag}(\mathbf{W}_{\rm{BB}}^{1},...,\mathbf{W}_{\rm{BB}}^{N_{R}^{\rm{group}}})\in {{\mathbb{C}}^{{{N}_{R}}\times N_{R}^{{}}}}$.

Further, by collecting ${\rm{\{ }}{{\bf{y}}_m}{\rm{\} }}_{m = 1}^{N_T^{{\rm{pilot}}}}$, we obtain an aggregate observation of the channel given by
\begin{shrinkeq}{-1ex}
\begin{align}
\begin{split}
\label{equ:Y}
\hspace{-4.6mm}
{\bf{Y}} & = {({\bf{W}}_{{\rm{RF}}}^{}{\bf{W}}_{{\rm{BB}}}^{})^H}{\bf{H}} {\kern 1pt} [{{\bf{s}}_1},...,{{\bf{s}}_{N_{T}^{{\rm{pilot}}}}}]{\rm{ + [}}{{{\bf{\bar n}}}_1},...,{{{\bf{\bar n}}}_{N_{T}^{{\rm{pilot}}}}}{\rm{]}}\\
& = {\bf{W}}_{{\rm{BB}}}^H{\bf{W}}_{{\rm{RF}}}^H{\bf{HF}}_{{\rm{RF}}}^{}{\bf{F}}_{{\rm{BB}}}^{} + {\bf{\bar N}} \text{,}
\end{split}
\end{align}
\end{shrinkeq}
where $\mathbf{Y}=[{{\mathbf{y}}_{1}},...,{{\mathbf{y}}_{N_{T}^{\rm{pilot}}}}] \in \mathbb{C}^{{N}_{R}\times {N}_{T}^{\rm{pilot}}}$,
$\mathbf{F}_{\rm{RF}}^{{}}=[\mathbf{F}_{\rm{RF}}^{1},$ $...,\mathbf{F}_{\rm{RF}}^{N_{T}^{\rm{group}}}] \in {{\mathbb{C}}^{{{N}_{T}}\times {{N}_{T}}}}$,
$\mathbf{F}_{\rm{BB}}^{{}}=\text{diag}(\mathbf{F}_{\rm{BB}}^{1},...,\mathbf{F}_{\rm{BB}}^{N_{T}^{\rm{group}}})\in {{\mathbb{C}}^{{{N}_{T}}\times N_{T}^{\rm{pilot}}}}$, $\mathbf{F}_{\rm{BB}}^{p} = [{\bf{f}}_{{\rm{BB}}}^{\frac{{(p - 1)N_T^{{\rm{pilot}}}}}{{N_T^{{\rm{group}}}}} + 1},...,{\bf{f}}_{{\rm{BB}}}^{\frac{{pN_T^{{\rm{pilot}}}}}{{N_T^{{\rm{group}}}}}}] \in {{\mathbb{C}}^{N_{T}^{\rm{RF}}\times \frac{N_{T}^{\rm{pilot}}}{N_{T}^{\rm{group}}}}}$, and ${{\bf{\bar N}}} = [{{{{\bf{\bar n}}}_1},...,{{{\bf{\bar n}}}_{N_{\rm{T}}^{{\rm{pilot}}}}}}] \in \mathbb{C}^{{N}_{R}\times {N}_{T}^{\rm{pilot}}}$.

{In a nutshell, in each Tx (Rx) group, $N_T^{\rm{RF}}$ ($N_R^{\rm{RF}}$) non-overlapping transmit (receive) training beams will be simultaneously formed by activating all RF chains at the transmitter (receiver).
With the pilot signals received by all Tx groups and all Rx groups, the FD angular-domain channels will be sounded.
For lens-array with simple ASN, our proposed CS-based approach takes into account the constraint imposed by practical ASN, so that the RF training beams are directional, which is different from the omni-directional RF training beams in \cite{peak,FD,DC,DL,image}.}
Note also that in each Tx group, the number of BB pilot signals is smaller than that of RF beams as we set $N_{T}^{{\rm{pilot}}} / N_{T}^{{\rm{group}}} < N_{T}^{{\rm{RF}}}$, i.e., $N_{T}^{{\rm{pilot}}} < N_T$, and the smaller $N_{T}^{{\rm{pilot}}} / N_{T}$ indicates the smaller training overhead.

{We note that a signal transmitted in mmWave band suffers from severely high path loss and blockage effect. Consequently, there exists only a few multipath components between the transmitter and the receiver in mmWave MIMO systems (i.e., $L \ll {N_T}$ or $L \ll {N_R}$). This property is often referred to as the {\it angular sparsity} \cite{Gao_CS, Gao_TSP, FGao}. Thanks to the angle-dependent energy focusing property of lens-array \cite{PDM, peak, FD}, along with the angular sparsity of mmWave channel, the channel matrix $\mathbf{H}$ can be readily modeled as a sparse matrix, indicating that only a small number of channel elements have the dominated channel energy.} By vectorizing $\mathbf{Y}$, therefore, a problem to estimate ${\bf{H}}$ from (\ref{equ:Y}) can be formulated as the sparse signal recovery problem as
\begin{shrinkeq}{-1ex}
\begin{align}
\label{equ:CSu}
{\bf{y}} = {\rm{vec(}}{\bf{Y}}{\rm{)}} &= ({\bf{F}}_{{\rm{RF}}}^T{\bf{F}}_{{\rm{BB}}}^T \otimes {\bf{W}}_{{\rm{BB}}}^H{\bf{W}}_{{\rm{RF}}}^H) {\rm{vec(}}{\bf{H}}{\rm{)}} + {\rm{vec(}}{\bf{\bar N}}{\rm{)}} \nonumber \\
& = {\bf{\Phi h}} + {\bf{\bar n}} \text{,}
\end{align}
\end{shrinkeq}
where $\mathbf{y}\in {{\mathbb{C}}^{N_{T}^{\rm{pilot}}N_{R}^{{}}\times 1}}$ is the vectorized received signal, $\mathbf{\Phi } = ({\bf{F}}_{{\rm{RF}}}^T{\bf{F}}_{{\rm{BB}}}^T \otimes {\bf{W}}_{{\rm{BB}}}^H{\bf{W}}_{{\rm{RF}}}^H) \in {{\mathbb{C}}^{N_{T}^{\rm{pilot}}N_{R}^{{}}\times N_{T}^{{}}N_{R}^{{}}}}$ is the measurement matrix, and $\mathbf{h}=\rm{vec(} \mathbf{H} \text{)} \in {{\mathbb{C}}^{\it{N}_{T}^{{}}N_{R}^{{}} \times \rm{1}}}$. Since $\mathbf{h}$ is a sparse vector, we can basically use any sparse signal recovery algorithms such as orthogonal matching pursuit (OMP) to efficiently estimate $\mathbf{h}$. Since $N_{T}^{\rm{pilot}}$ is in general significantly smaller than $N_{T}^{{}}$, pilot overhead of our approach is much lower than that required by conventional approaches such as least square (LS) estimation technique \cite{peak, FD}. {Also note that the measurement matrix $\mathbf{\Phi }$ in (\ref{equ:CSu}) is dedicated to the proposed framework of pilot training, which is essential to the design of BB pilot in Section III-C.}

\vspace*{-6mm}
\subsection{Redundant Dictionary Design to Sparsify Channels}
{The accuracy of CS-based CE depends heavily on the sparsity of $\mathbf{h}$ in (\ref{equ:CSu}) \cite{BShim}. However, the power leakage caused by the mismatch between continuous AoAs/AoDs and discrete dictionary with limited resolution may weaken the sparsity of $\mathbf{h}$ \cite{DC}. To mitigate this behaviour, we design a redundant dictionary by quantizing both the virtually vertical and horizontal angles with a finer resolution, where the sets of the quantized virtual angles can be expressed as
\begin{shrinkeq}{-1ex}
\begin{align}
\label{equ:sets}
\begin{split}
{\cal A}_{v} & = \{ {\theta _g} | {\sin {\theta_g}} =  -1+(2g - 1) / G_{v},g = 1,...,G_{v} \},\\
{\cal A}_{h} & = \{ {\varphi _g} | {\sin {\varphi_g}} =  -1+(2g - 1)/ G_{h},g = 1,...,G_{h} \} \text{.}
\end{split}
\end{align}
\end{shrinkeq}}Here ${{\mathcal{A}}_{v}}$ and ${{\mathcal{A}}_{h}}$ are the sets of quantized vertical and horizontal angles, respectively, and ${G_{v}}{G_{h}} \gg \max (N_{R}^{},N_{T}^{})$. Under this setting, the channel matrix can be expressed as
\begin{shrinkeq}{-1ex}
\begin{align}
\label{equ:Ha}
\hspace{-4.6mm}
\mathbf{H}={{\mathbf{A}}_{R}}{{\mathbf{H}}_{a}}\mathbf{A}_{T}^{H}+\mathbf{E} \text{,}
\end{align}
\end{shrinkeq}
where
\begin{shrinkeq}{-1ex}
\begin{align*}
\hspace{-4.6mm}
{{\mathbf{A}}_{R}} & = [{{\mathbf{a}}_{R}}({{\theta }_{1}},{{\varphi }_{1}}),...,{{\mathbf{a}}_{R}}({{\theta }_{1}},{{\varphi }_{{{G}_{h}}}}),...,{{\mathbf{a}}_{R}}({{\theta }_{{{G}_{v}}}},{{\varphi }_{{{G}_{h}}}})] \text{,} \\
{{\mathbf{A}}_{T}} & = [{{\mathbf{a}}_{T}}({{\theta }_{1}},{{\varphi }_{1}}),...,{{\mathbf{a}}_{T}}({{\theta }_{1}},{{\varphi }_{{{G}_{h}}}}),...,{{\mathbf{a}}_{T}}({{\theta }_{{{G}_{v}}}},{{\varphi }_{{{G}_{h}}}})]
\end{align*}
\end{shrinkeq}
are the dedicated redundant dictionaries for lens-array, ${{\mathbf{H}}_{a}}\in {{\mathbb{C}}^{{{G}_{v}}{{G}_{h}}\times {{G}_{v}}{{G}_{h}}}}$ is the $L_a$-sparse channel approximation in the quantized virtual angular domain, $\mathbf{E}\in {{\mathbb{C}}^{{{N}_{R}}\times {{N}_{T}}}}$ is the quantization error matrix treated as a random noise. Note that the redundant dictionary design in (\ref{equ:Ha}) is tailored for FD lens-array according to (\ref{sinc}), which is essentially different from the dictionary design for phased uniform linear array in \cite{korean}.

By substituting (\ref{equ:Ha}) into (\ref{equ:CSu}), we have
\begin{shrinkeq}{-1ex}
\begin{align}
\begin{split}
\hspace{-4.6mm}
\label{equ:CSr}
{\bf{y}}{\kern 1pt} = {\bf{\Phi }} {\rm{vec(}}{{\bf{A}}_{\rm{R}}}{{\bf{H}}_a}{\bf{A}}_{\rm{T}}^H + {\bf{E}}{\rm{)}} + {\bf{\bar n}} = {\bf{\Phi \Psi }}{{\bf{h}}_a} + {\bf{n}} \text{,}
\end{split}
\end{align}
\end{shrinkeq}
where $\mathbf{\Psi } = {\bf{A}}_{\rm{T}}^ *  \otimes {{\bf{A}}_{\rm{R}}} \in {{\mathbb{C}}^{N_{T}^{{}}N_{R}^{{}}\times {{({{G}_{v}}{{G}_{h}})}^{2}}}}$ is the redundant dictionary matrix, ${{\mathbf{h}}_{a}}=\text{vec}({{\mathbf{H}}_{a}})\in {{\mathbb{C}}^{{{({{G}_{v}}{{G}_{h}})}^{2}}\times 1}}$ is the enhanced $L_a$-sparse channel vector after the transformation by $\mathbf{\Psi }$, and $\mathbf{n}={\bf{\Phi }} {\rm{vec}}({\bf{E}}) + {\bf{\bar n}}$ is the effective noise vector. Usually, the quantization error $\mathbf{E}$ and the degradation of the sparsity of $\mathbf{H}$ can be mitigated by increasing ${{G}_{v}}$ and ${{G}_{h}}$. Thus, to obtain a better performance, we first use (\ref{equ:CSr}) to estimate $\mathbf{h}_a$ and then get the estimate of $\mathbf{H}$ using (\ref{equ:Ha}).

\begin{table*}[t]
\vspace{-8mm}
\renewcommand\arraystretch{1.0}
\centering
\caption{Computational Complexity of Two Schemes in Each Iteration}
\label{TAB1}
\vspace{-4mm}
\begin{center}
\begin{tabular}{c|l!{\vrule width1pt}l}
\Xhline{1.2pt}
  & \multicolumn{1}{c!{\vrule width1pt}}{\textbf{Proposed Scheme with OMP}} &
	\multicolumn{1}{c}{\textbf{DC-based Support Detection Scheme} \cite{DC}} \\
\Xhline{1pt}
	\multirow{2}{*}{\makecell*[c]{Correlation}} & \multirow{2}{*}{$\textsf{O} \big( N_R N_T^{{\rm{pilot}}}{G_v}{G_h} \big)$} & $\textsf{O} \big( ({N_h} - 1)(8N_h^3 + 8N_h^2 + 2{N_h}N_T^{{\rm{pilot}}} + 2{N_h})$ \\ & & \qquad \ \ \ $+({N_v} - 1)(8N_v^3 + 8N_v^2 + 2{N_v}N_T^{{\rm{pilot}}} + 2{N_v}) \big)$ \\

\hline
	Project subspace & $\textsf{O} \big( {i^3} + 2 N_R N_T^{{\rm{pilot}}}{i^2} + N_R N_T^{{\rm{pilot}}}i \big)$ & $\textsf{O}\big({J^3} + 2{J^2}N_T^{{\rm{pilot}}} + J N_T^{{\rm{pilot}}} \big)$ \\
\hline
 Update residual & $\textsf{O}\big( N_R N_T^{{\rm{pilot}}}i \big)$ & $\textsf{O}\big( N_T^{{\rm{pilot}}}J \big)$ \\
\hline
 Stop criteria & $\textsf{O}\big( N_R N_T^{{\rm{pilot}}} \big)$ & N/A \\

\Xhline{1.2pt}
\end{tabular}
\end{center}
\vspace{-4mm}
\end{table*}

\vspace*{-5mm}
\subsection{Pilot Signals Design Based on CS Theory}
{To obtain a measurement matrix with a good RIP, many CS-based CE techniques employ the pilot signals randomized by the phase-shifters in the RF part \cite{DC,image,DL}. To do so, an extra PSN, as shown in Fig. 1(a), is required. Nevertheless, designing pilot signals for lens-array with simple ASN to achieve the i.i.d. entries of measurement matrix with good RIP is not possible due to the hardware constraint resulted from ASN.
This motivates us to design the pilot signals constructing the measurement matrix $\mathbf{\Phi}$ in (\ref{equ:CSr}), based on the more tractable total mutual coherence minimization criterion \cite{BShim, 22, Elad, Gao_CS} for reliable sparse CE.} Since we adopt the simple ASN, the RF parts of the pilot signal can be expressed as
\begin{shrinkeq}{-1ex}
\begin{align}
\hspace{-4.6mm}
\label{equ:RFmtx}
\mathbf{F}_{\rm{RF}}^{{}}={{\mathbf{\tilde{I}}}_{{{N}_{T}}}} \text{,} \mathbf{W}_{\rm{RF}}^{{}}={{\mathbf{\tilde{I}}}_{{{N}_{R}}}} \text{,}
\end{align}
\end{shrinkeq}
where ${{\mathbf{\tilde{I}}}_{N}}$ is the $N\times N$ identity matrix after the random permutation among its columns, and the elements ``1'' and ``0'' denote switching on and off, respectively.
Note that we randomly permute the columns of $\mathbf{F}_{\rm{RF}}^{{}}$ and $\mathbf{W}_{\rm{RF}}^{{}}$ to improve the CE performance.

Moreover, given $\mathbf{F}_{\rm{RF}}^{{}}$ and $\mathbf{W}_{\rm{RF}}^{{}}$, we minimize the total mutual coherence \cite{BShim, Elad, 22, Gao_CS} of the matrix $\mathbf{\Phi}\mathbf{\Psi}$ by formulating the design problem of $\mathbf{F}_{\rm{BB}}^{{}}$ and $\mathbf{W}_{\rm{BB}}^{{}}$ as
\begin{shrinkeq}{-1ex}
\begin{align}
\begin{split}
\hspace{-4.6mm}
\label{equ:opt}
&(\mathbf{F}^{\star}_{\rm{BB}},\mathbf{W}^{\star}_{\rm{BB}})=\underset{\mathbf{F}_{\rm{BB}}^{{}},\mathbf{W}_{\rm{BB}}^{{}}}{\mathop{\arg \min }}\,\mu^t (\mathbf{\Phi \Psi }) \\
&\text{s.t.} \left\| {{{{\bf{F}}_{{\rm{BB}}}}}} \right\|_F^2 = N_T^{\rm{pilot}} \quad \text{and} \quad \left\| {{{{\bf{W}}_{{\rm{BB}}}}}} \right\|_F^2 = N_R \text{,}
\end{split}
\end{align}
\end{shrinkeq}
where $\mu^t (\mathbf{\Phi \Psi })\triangleq \sum_{m\ne n}{\left| [\mathbf{\Phi \Psi }]_{:,m}^{H}[\mathbf{\Phi \Psi }]_{:,n}^{{}} \right| ^2 }$ is the total mutual coherence of $\mathbf{\Phi \Psi }$, and we assume that the transmit power is normalized $( \left\| {{{[{{\bf{F}}_{{\rm{BB}}}}]}_{:,i}}} \right\|_2^2 = 1 , 1 \le i \le N_T^{\rm{pilot}})$ and $\left\| {{{[{{\bf{W}}_{{\rm{BB}}}}]}_{:,j}}} \right\|_2^2 = 1,1 \le j \le N_R$ for ease of analysis, respectively. According to \cite{korean}, $\mu^t (\mathbf{\Phi \Psi })$ satisfies the following inequality
\begin{shrinkeq}{-1ex}
\begin{equation}
\mu^t ({\bf{\Phi \Psi }}) \le \mu^t ({\bf{F}}_{{\rm{BB}}}^T{\bf{F}}_{{\rm{RF}}}^T{\bf{A}}_T^*) \mu^t ({\bf{W}}_{{\rm{BB}}}^H{\bf{W}}_{{\rm{RF}}}^H{{\bf{A}}_R}){\text{,}}
\end{equation}
\end{shrinkeq}
which sheds light on how we decouple the problem (\ref{equ:opt}) to avoid the difficulty in joint optimization. To be specific, we minimize $\mu^t ({\bf{F}}_{{\rm{BB}}}^T{\bf{F}}_{{\rm{RF}}}^T{\bf{A}}_T^*)$ and $\mu^t ({\bf{W}}_{{\rm{BB}}}^H{\bf{W}}_{{\rm{RF}}}^H{{\bf{A}}_R})$ over $\mathbf{F}_{\rm{BB}}$ and $\mathbf{W}_{\rm{BB}}$, respectively. Taking $\mu^t ({\bf{F}}_{{\rm{BB}}}^T{\bf{F}}_{{\rm{RF}}}^T{\bf{A}}_T^*)$ as an example, we have
\begin{align*}
& \mu^t ({\bf{F}}_{{\rm{BB}}}^T{\bf{F}}_{{\rm{RF}}}^T{\bf{A}}_T^*) \mathop  = \limits^{(a)} \left\| {{\bf{F}}_a^H{{\bf{F}}_a} - {{\bf{I}}_{{G_e}{G_a}}}} \right\|_F^2 \nonumber \\
& = {\rm{Tr(}}{\bf{F}}_a^H{{\bf{F}}_a}{\bf{F}}_a^H{{\bf{F}}_a} - 2{\bf{F}}_a^H{{\bf{F}}_a} + {{\bf{I}}_{{G_e}{G_a}}}{\rm{)}} \nonumber \\
& = {\rm{Tr(}}{{\bf{F}}_a}{\bf{F}}_a^H{{\bf{F}}_a}{\bf{F}}_a^H - 2{{\bf{F}}_a}{\bf{F}}_a^H + {{\bf{I}}_{N_T^{{\rm{pilot}}}}}{\rm{) + }}{G_e}{G_a} - N_T^{{\rm{pilot}}}\nonumber \\
& = \left\| {{{\bf{F}}_a}{\bf{F}}_a^H - {{\bf{I}}_{N_T^{{\rm{pilot}}}}}} \right\|_F^2{\rm{ + }}{G_e}{G_a} - N_T^{{\rm{pilot}}}\nonumber
\end{align*}
\begin{align}
& = \left\| {{\bf{F}}_{{\rm{BB}}}^T{\bf{F}}_{{\rm{RF}}}^T{\bf{A}}_T^*{\bf{A}}_T^T{\bf{F}}_{{\rm{RF}}}^*{\bf{F}}_{{\rm{BB}}}^* - {{\bf{I}}_{N_T^{{\rm{pilot}}}}}} \right\|_F^2{\rm{ + }}{G_e}{G_a} - N_T^{{\rm{pilot}}}\nonumber \\
& \mathop  \approx \limits^{(b)} \left\| {{c_T}{\bf{F}}_{{\rm{BB}}}^T{\bf{F}}_{{\rm{BB}}}^* - {{\bf{I}}_{N_T^{{\rm{pilot}}}}}} \right\|_F^2{\rm{ + }}{G_e}{G_a} - N_T^{{\rm{pilot}}} \nonumber \\
& \mathop  = \limits^{(c)} \sum\nolimits_{p = 1}^{N_T^{{\rm{group}}}} {\left\| {{c_T}{{({\bf{F}}_{{\rm{BB}}}^p)}^T}{{({\bf{F}}_{{\rm{BB}}}^p)}^*} - {{\bf{I}}_{N_T^{{\rm{pilot}}}/N_T^{{\rm{group}}}}}} \right\|_F^2} \nonumber \\
& \qquad \qquad \qquad \qquad \qquad +{G_e}{G_a} - N_T^{{\rm{pilot}}} \text{,}
\label{equ:myopt}
\end{align}
where (a) is based on ${{\bf{F}}_a} = {\bf{F}}_{{\rm{BB}}}^T{\bf{F}}_{{\rm{RF}}}^T{\bf{A}}_T^*$ and the normalized ${\ell _2}$-norm assumption for each column of ${{\bf{F}}_a}$, (b) follows from $\mathbf{F}_{\rm{RF}}^{{H}}\mathbf{F}_{\rm{RF}}=\mathbf{I}_{N_T}$ and ${{\mathbf{A}}_{T}}\mathbf{A}_{T}^{H}\approx {{c}_{T}}{{\mathbf{I}}_{N_{T}^{{}}}}${\footnote{This approximation is empirically reasonable for the dictionary matrix $\mathbf{A}_T$ and we can use the metric ${{{\varepsilon }_{T}}=\left\| {{\mathbf{A}}_{T}}\mathbf{A}_{T}^{H}-{{c}_{T}}{{\mathbf{I}}_{N_{T}^{{}}}} \right\|_{F}^{2}}/{\left\| {{\mathbf{A}}_{T}}\mathbf{A}_{T}^{H} \right\|_{F}^{2}}$ with ${{c}_{T}}=\text{Tr}({{\mathbf{A}}_{T}}\mathbf{A}_{T}^{H})/{{N}_{T}}$ to justify it. In our simulations, we calculate that the value of ${\varepsilon }_{T}$ will be smaller than $0.2$, which is sufficiently small to ensure the validity of this approximation.}, and (c) is due to $\mathbf{F}_{\rm{BB}}^{{}}=\text{diag}(\mathbf{F}_{\rm{BB}}^{1},...,\mathbf{F}_{\rm{BB}}^{N_{T}^{\rm{group}}})$. One can readily observe that columns from $\mathbf{F}_{\rm{BB}}^{p}, \forall p \in \{1,...,N_{T}^{\rm{group}}\}$, should be mutually orthogonal to minimize (\ref{equ:myopt}). Therefore, the solution $\mathbf{F}^{\star}_{\rm{BB}}$ minimizing (\ref{equ:myopt}) can be expressed as
\begin{shrinkeq}{-1ex}
\begin{align}
\hspace{-4.6mm}
\label{equ:mysoluF}
\mathbf{F}^{\star}_{\rm{BB}} = \text{diag} (\mathbf{F}_{\rm{BB}}^{1 \star},...,\mathbf{F}_{\rm{BB}}^{N_{T}^{\rm{group}} \star})\text{,}
\end{align}
\end{shrinkeq}
where ${\bf{F}}_{{\rm{BB}}}^{p \star} (1 \le p \le {N_{T}^{\rm{group}}})$ are the matrices satisfying ${{({\bf{F}}_{{\rm{BB}}}^{p \star})^H}{\bf{F}}_{{\rm{BB}}}^{p \star} = {\bf{I}}_{N_T^{{\rm{pilot}}}/N_T^{{\rm{group}}}}}$. In our simulations, we use the specific solution $\mathbf{F}_{\rm{BB}}^{p \star} = {{\bf{U}}_{1:N_T^{{\rm{pilot}}}/N_T^{{\rm{group}}}}}$, where ${{\mathbf{U}}}$ is the $N_T^{{\rm{RF}}} \times N_T^{{\rm{RF}}}$ discrete Fourier transformation (DFT) matrix and the notation ${{\bf{U}}_{1:N_T^{{\rm{pilot}}}/N_T^{{\rm{group}}}}}$ denotes the submatrix of ${{\mathbf{U}}}$ constructed from the first $\frac{N_{T}^{\rm{pilot}}}{N_{T}^{\rm{group}}}$ columns of ${{\mathbf{U}}}$. Similarly, the obtained solution for $\mathbf{W}_{\rm{BB}}$ is given by
\begin{shrinkeq}{-1ex}
\begin{align}
\hspace{-4.6mm}
\label{equ:mysoluW}
\mathbf{W}^{\star}_{\rm{BB}} = \text{diag} (\mathbf{W}_{\rm{BB}}^{1 \star},...,\mathbf{W}_{\rm{BB}}^{N_{R}^{\rm{group}} \star})\text{,}
\end{align}
\end{shrinkeq}
where $\mathbf{W}_{\rm{BB}}^{n \star} (1 \le n \le N_R^{\rm{group}})$ can be arbitrary unitary matrices. In our simulations, we set $\mathbf{W}_{\rm{BB}}^{n \star}$ as the DFT matrix.

\begin{figure*}[!tp]
\vspace{-5mm}
\captionsetup{font={footnotesize, color = {black}}, name = {Fig.}, labelsep = period}
\captionsetup[subfigure]{singlelinecheck = on, justification = raggedright, font={footnotesize}}
\centering
\subfloat{
\label{3a}
\begin{minipage}[t]{0.4\linewidth}
\centering
\includegraphics[width=2.6in]{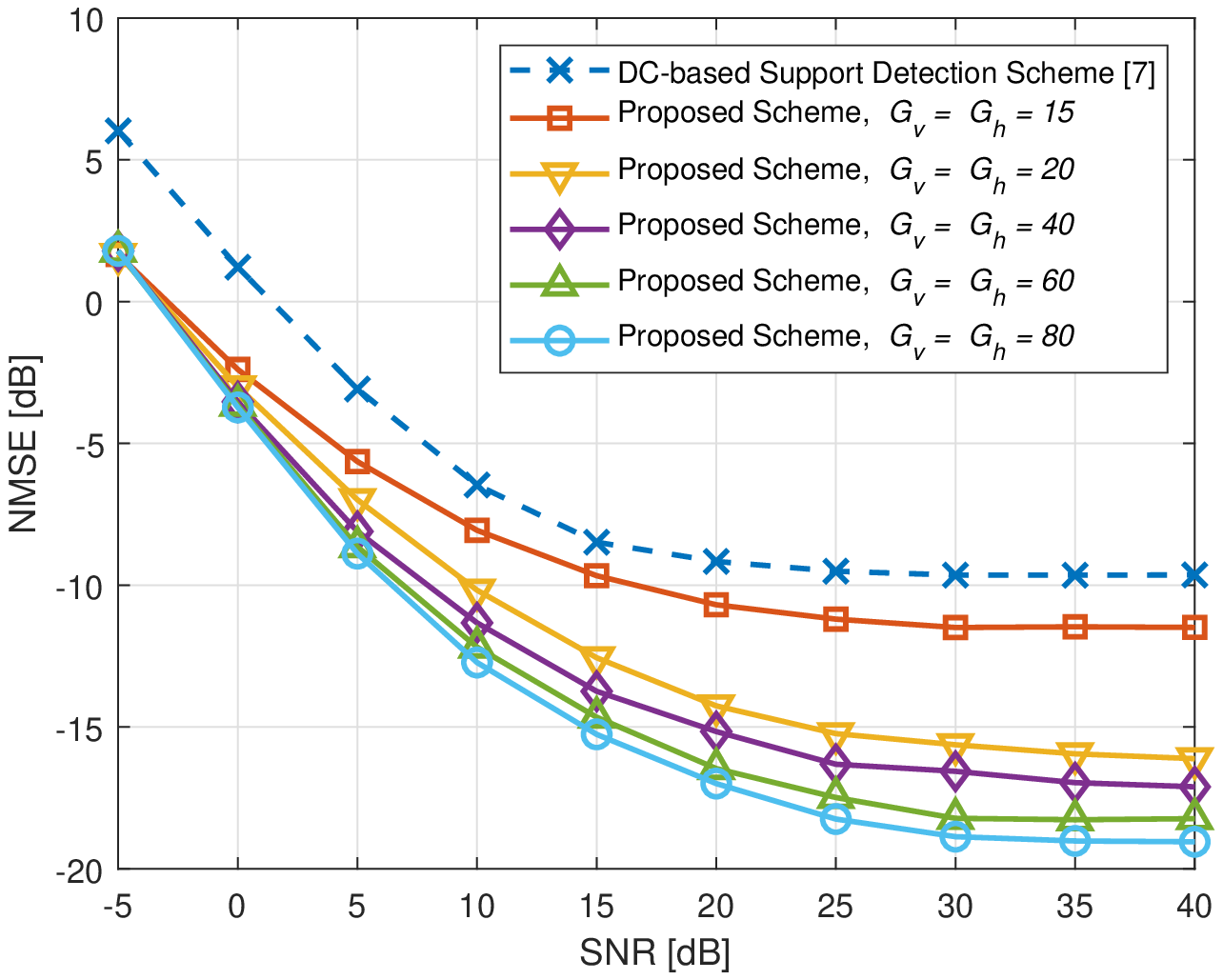}
\centerline{(a)}
\end{minipage}
}
\subfloat{
\label{3b}
\begin{minipage}[t]{0.4\linewidth}
\centering
\includegraphics[width = 2.6in]{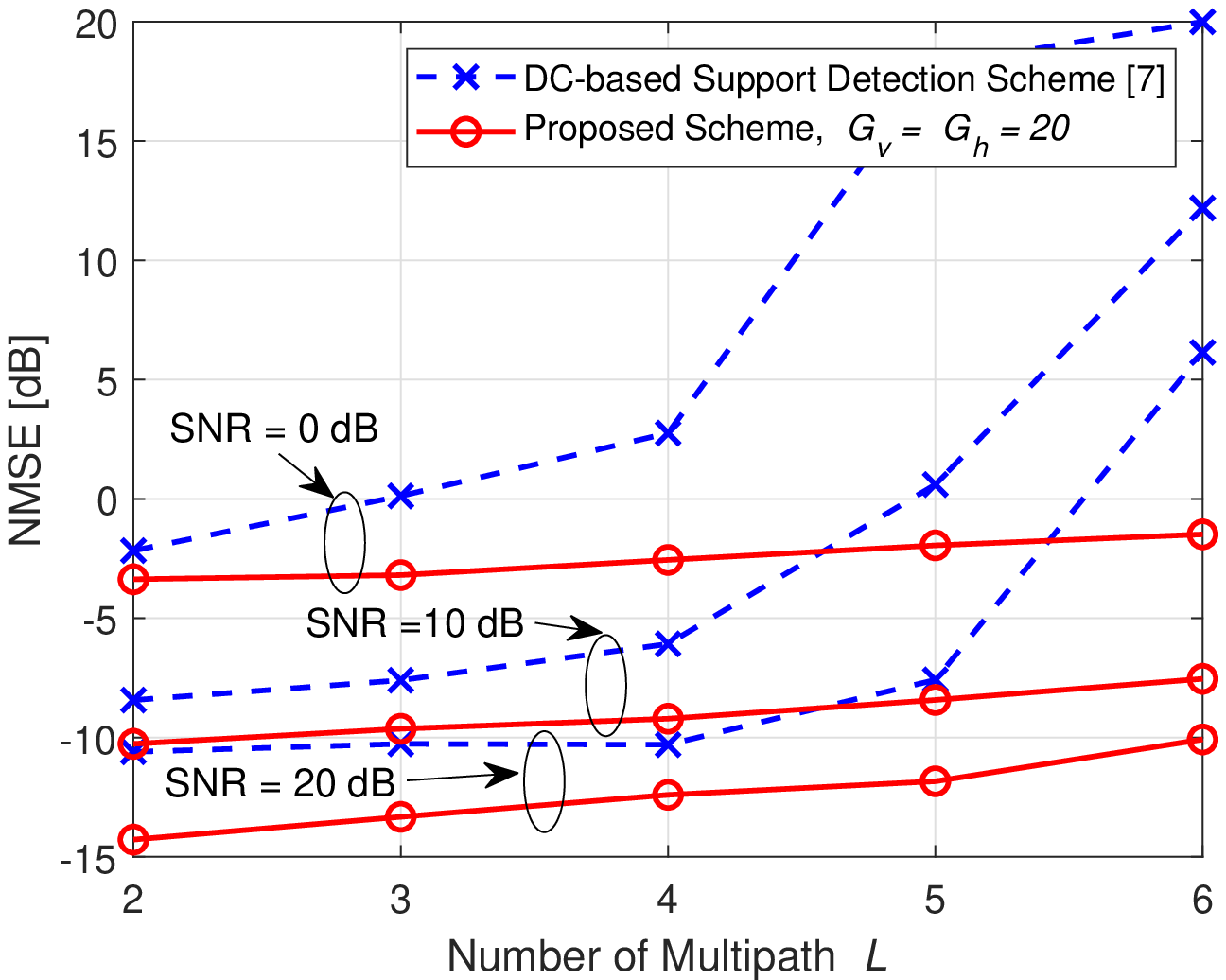}
\centerline{(b)}
\end{minipage}
}
\vspace{1.0mm}
\caption{Performance comparsion between the proposed scheme and the DC-based support detection scheme with ${{N}_{R}}=N_{R}^{\rm{RF}}=1$, ${{N}_{T}}=128$, $N_{T}^{\rm{RF}}=4$ and ${{N}_{T}^{\rm pilot}}=64$.}
\vspace{-3.0mm}
\end{figure*}

\begin{figure*}[!tp]
\captionsetup{font={footnotesize}, name = {Fig.}, labelsep = period}
\captionsetup[subfigure]{singlelinecheck = on, justification = raggedright, font={footnotesize}}
\centering
\subfloat{
\label{4a}
\begin{minipage}[t]{0.4\linewidth}
\centering
\includegraphics[width=2.6in]{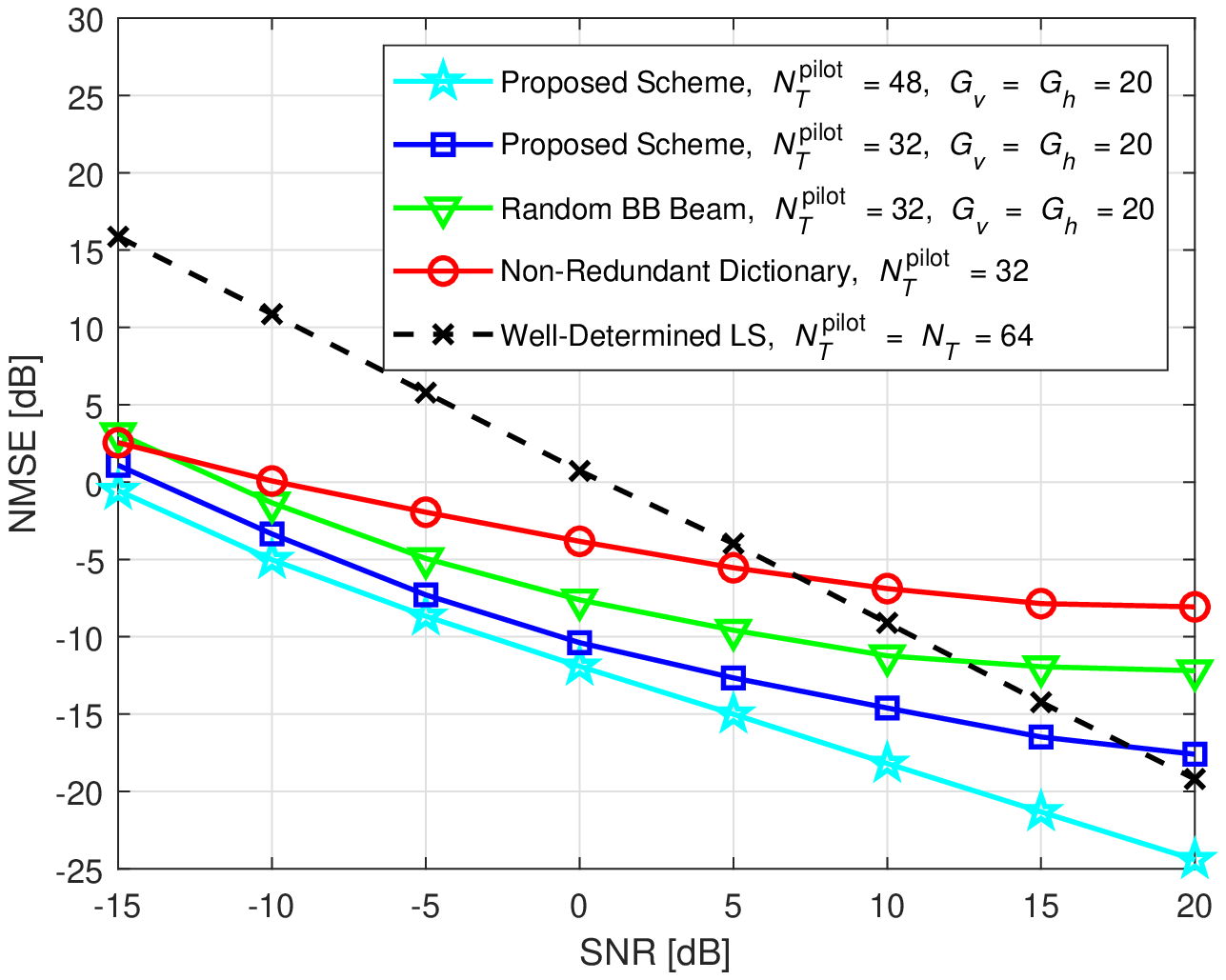}
\centerline{(a)}
\end{minipage}
}
\subfloat{
\label{4b}
\begin{minipage}[t]{0.4\linewidth}
\centering
\includegraphics[width = 2.6in]{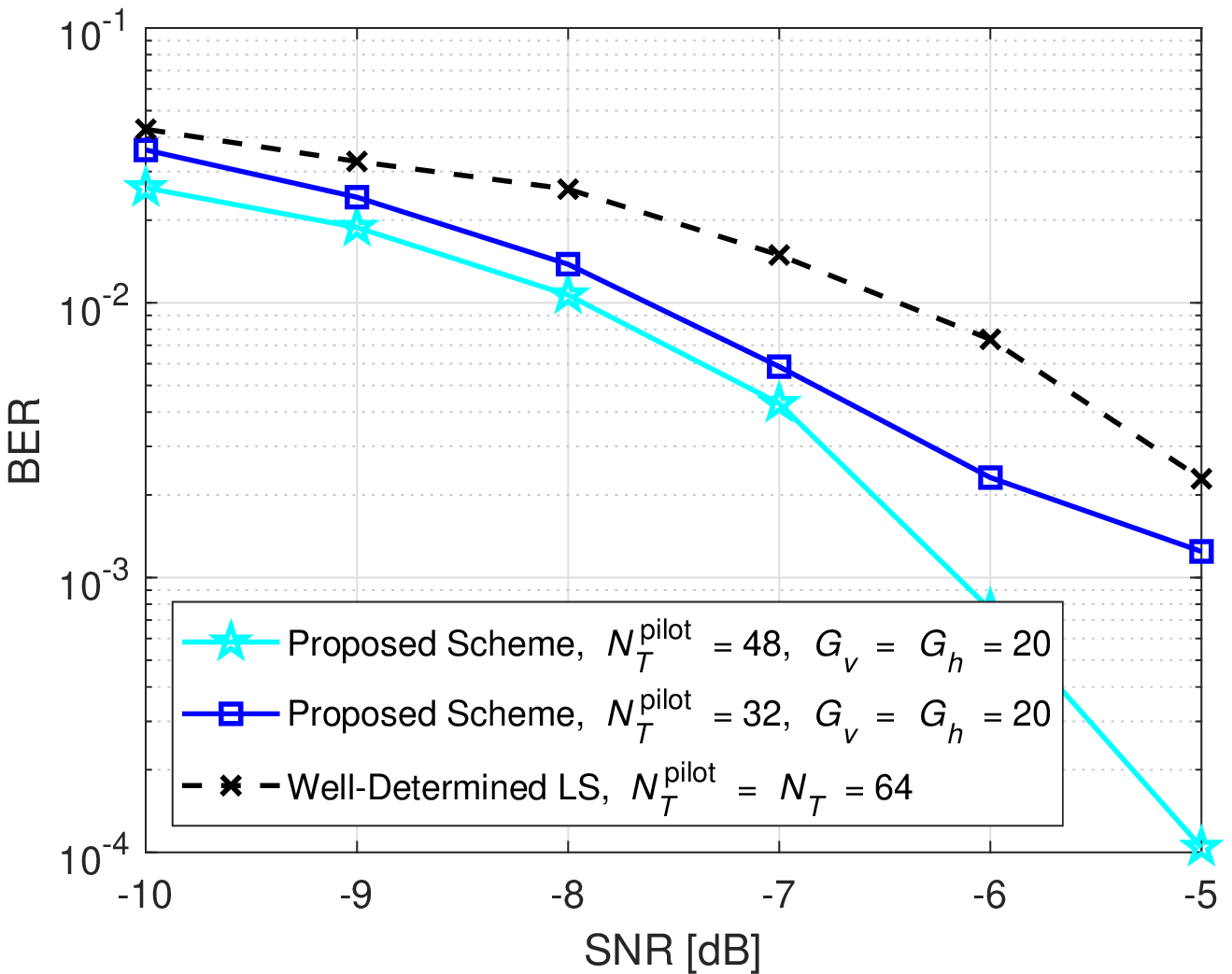}
\centerline{(b)}
\end{minipage}
}
\vspace{1.0mm}
\caption{Simulation results of different schemes with ${{N}_{T}}={{N}_{R}}=64$ and $N_{T}^{\rm{RF}}=N_{R}^{\rm{RF}}=4$: (a) NMSE; (b) BER.}
\vspace{-3.0mm}
\end{figure*}

\vspace*{-3mm}
\subsection{Computational Complexity Analysis}
In this subsection, we focus on the computational complexity of the proposed scheme.
Since the CE problem has been formulated as the sparse signal recovery problem in Section III-B, various off-the-shelf CS algorithms can be used to estimate the channels for the proposed scheme. Clearly, the computational complexity of the proposed scheme heavily depends on the adopted sparse signal recovery algorithms. In this paper, we employ the well-known OMP algorithm \cite{korean,BShim} to solve $\mathbf{h}_a$ in (\ref{equ:CSr}). The computational complexity of our scheme is summarized in the left column of Table \ref{TAB1}. In Table \ref{TAB1}, the notation $\textsf{O}(N)$ stands for ``on order of $N$'', and the index of iteration in OMP algorithm is denoted by $i$. In essence, the OMP algorithm consists of four major steps: correlation, project subspace, residual update, and stop criteria identification, and the computational complexity of each step is summarized in the table. Taking a typical system configuration with $N_R = 1$, $N_T^{\rm pilot} = 32$ and $G_v \times G_h = 20 \times 20$ as an example, we can observe that the most significant computational complexity burden comes from the step of correlation due to the large $G_v$ and $G_h$. However, the computational complexity of other steps, especially for the step of project subspace requiring matrix inversion operation, is irrelevant to $G_v$ and $G_h$, which indicates that the complexity of the proposed scheme is acceptable even though large $G_v$ and $G_h$ are choosen for better resolution of the redundant dictionary.

To compare the proposed scheme with existing techniques, we consider the state-of-the-art dual crossing (DC)-based support detection algorithm \cite{DC}. The DC-based support detection algorithm is an extension of OMP algorithm, and its computational complexity is provided in the right column of Table \ref{TAB1}. In \cite{DC}, ${N_v} \times {N_h} = {N_T}$ is the geometric size of transmit lens-array, the reciever has one antenna with $N_R = 1$, and $J$ is a pre-defined parameter. Note that $J$ can be very large compared with the number of multipath $L$ (e.g., $J$ was set to $64$ when $L=3$ in \cite{DC}). Therefore, the DC-based support detection algorithm suffers from high computational complexity due to high-dimensional matrix inversion operation in project subspace step. By contrast, although the computational complexity of the proposed scheme increases with $G_v$ and $G_h$, it can provide a more robust performance and a trade-off between the performance and the complexity, which will be detailed in the next section.

\vspace*{-2mm}
\section{Simulation Results}
\vspace*{-1mm}
In the simulation, for the channel model, we set the carrier frequency to $30$ GHz \cite{Gao}, ${{g }_{l}}\sim \mathcal{C}\mathcal{N}(0,1)$, and the AoAs/AoDs $\theta _{R}^{l}$, $\varphi _{R}^{l}$, $\theta _{T}^{l}$, $\varphi _{T}^{l}$ follow a uniform distribution ${\cal U}[-\pi /2,\pi /2]$.
First, we compare the performance of the proposed scheme and the DC-based support detection scheme \cite{DC}. For fairness, we consider a downlink system with a single-antenna user, i.e. ${{N}_{R}}={{N}_R^{\rm RF}}=1$, consistent with those in \cite{DC}. We set $\tilde{D}_{T}^{h}\times \tilde{D}_{T}^{v}=6.4 \times 6.4$ with ${{N}_{T}}=128$ according to (1), which is equivalent to an ${N_v} \times {N_h} = 8 \times 16$ uniform rectangular array in \cite{DC}, and $N_{T}^{\rm{RF}}=4$. Note that the DC-based support detection scheme requires the bulky full-connected PSN, so that it will cause a prohibitively large hardware cost.

In Fig. 3(a), we plot the normalized mean square error (NMSE) performance of the DC-based support detection scheme and the proposed scheme with different $\{G_v,G_h\}$ as a function of signal-to-noise ratio (SNR), where $L=3$ is considered. It can been seen that even with a simple ASN, the proposed scheme outperforms the DC-based support detection scheme when the appropriate parameters $\{G_v,G_h\}$ are chosen. {We also observe that the performance of the proposed scheme improves with $\{G_v,G_h\}$, which trades off the NMSE performance with the computational complexity. Moreover, when $\{G_v, G_h\}$ is larger than $20$, the performance improvement is minor, but at the cost of prohibitive computational complexity. For this reason, we set $G_v = G_h = 20$ in our experiments.}

We further investigate the robustness of different schemes as a function of the number of multipath $L$ in Fig. 3(b). Note that when $L$ increases, the power leakage becomes severe and the structured sparsity patterns of different paths leveraged in the DC-based support detection scheme are destroyed. As a result, we observe from Fig. 3(b) that the performance of the DC-based support detection scheme is degraded when $L$ increases. However, by leveraging the enhanced sparsity benefited from the designed redundant dictionary, the proposed scheme can effectively estimate the channels with more multipath components.

Another drawback of the DC-based support detection scheme in \cite{DC} is that it only considers the systems with single-antenna users, while the proposed scheme can be directly applied to the system with lens-array at both the transmitter and the receiver. We consider such a more general scenario and investigate the performance of the proposed scheme. In simulations, we consider $\tilde{D}_{T}^{h}\times \tilde{D}_{T}^{v}=\tilde{D}_{R}^{h}\times \tilde{D}_{R}^{v}=4.7 \times 4.7$ with ${{N}_{T}}={{N}_{R}}=64$ according to (\ref{equ:Nt}), $N_{T}^{\rm{RF}}=N_{R}^{\rm{RF}}=4$ (i.e., $N_{T}^{\rm{group}}=N_{R}^{\rm{group}}=16$) and ${{G}_{v}}={{G}_{h}}=20$.

\color{black}
For comparison, we also investigate the following schemes: 1) the proposed scheme with random BB pilot signals \cite{Gao}; 2) the proposed scheme without using redundant dictionary, and 3) the conventional well-determined LS estimator based on the beam training, i.e., using LS to estimate $\mathbf{h}$ from (\ref{equ:CSu}) when $N_T^{\rm{pilot}}=N_T$, $\mathbf{F}_{\rm{BB}} = \mathbf{I}_{N_T}$ and $\mathbf{W}_{\rm{BB}} = \mathbf{I}_{N_R}$.

In Fig. 4(a), we plot the NMSE performance of different schemes as a function of SNR. We observe that the NMSE of the proposed scheme improves with $N_T^{\rm{pilot}}$. {We also observe that the gain of the proposed BB pilot signal design and the redundant dictionary design is considerable within a wide range of SNR.}
{For example, when $N_T^{\rm{pilot}}=32$ and SNR $=0$ dB, the proposed scheme has more than $10$ dB gain over the well-determined LS scheme with only half the pilot resources.}
Note that for beam training based LS estimator, $\bf{\Phi}$ in (\ref{equ:CSu}) becomes a unitary matrix, so the NMSE performance of LS estimator achieves the minimum of Cram\'{e}r-Rao lower bound of linear estimators. {However, by leveraging the sparsity of mmWave channels, the proposed CS-based CE scheme outperforms the LS estimator even with much smaller number of pilots, especially at low SNR. Considering that the SNR is usually low in most mmWave systems at the CE stage, the proposed scheme is effective in estimating the practical channels for mmWave FD-MIMO with lens-array.}

We further compare the bit-error-rate (BER) performance of the proposed CE scheme and well-determined LS approach. Based on the estimated channel, we apply an energy-based antenna selection scheme in \cite{peak} for the data transmission. We consider the 64-QAM modulation, SVD precoding, and turbo channel coding with 1/3 code rate. From Fig. 4(b) we see that when compared to the well-determined LS scheme, the proposed scheme achieves improved BER performance with a reduced pilot overhead.
{When BER is $10^{-2}$, for example, the proposed scheme with $N_T^{\rm{pilot}}=32$ achieves about $1$ dB gain over the LS scheme, while the pilot overhead is reduced by $50 \%$.}

\vspace*{-2mm}
\color{black}
\section{Conclusions}
In this paper, we proposed a CS-based CE scheme for mmWave FD-MIMO with lens-array, which sheds light on the application of CS techniques to lens-array using simple and practical ASN.
Specifically, we first proposed a framework of pilot training based on CS under the constraint imposed by ASN.
Then, we designed the dedicated redundant dictionary tailored for FD lens-array.
We also designed the transmit/receive pilot signals for improved CE performance. In particular, the BB pilot signals are designed to minimize the total mutual coherence of the measurement matrix. From the simulation results, we demonstrated the effectiveness of the proposed CE technique.

\end{document}